\documentclass[12pt, amsmath, preprint]{iopart}
\usepackage{amssymb}
\usepackage{amsfonts}
\usepackage{iopams}
\usepackage{graphicx}
\begin{document}

\title{ Detailed study of dissipative quantum dynamics of K$_2$
attached to helium nanodroplets}

\author{Martin Schlesinger, Walter T. Strunz}
\address{Institut f\"ur Theoretische Physik, Technische Universit\"at
Dresden, D-01062 Dresden, Germany}
\ead{Martin.Schlesinger@tu-dresden.de}
\date{\today}
\begin{abstract}
We thoroughly 
investigate vibrational quantum dynamics of dimers attached to He droplets 
motivated by recent measurements with 
K$_2$ \cite{Claas_1151_2006}. 
For those femtosecond pump-probe experiments, 
crucial observed features are not reproduced 
by gas phase calculations but agreement is found 
using a description based on dissipative quantum dynamics, 
as briefly shown in \cite{Schle_245_2010}. 
Here we present a detailed study of the influence of 
 possible effects induced by the droplet. 
 The helium 
 droplet causes electronic decoherence, 
 shifts of potential surfaces, and relaxation of 
 wave packets in attached dimers. 
 Moreover, a realistic description of (stochastic) desorption 
 of dimers off the droplet needs to be taken into account. 
 Step by step we include 
 and study the importance of these effects 
 in our full quantum calculation. 
 This allows us to reproduce and explain all major experimental findings. 
 We find that desorption is fast and occurs already 
 within $2-10$ ps after electronic excitation. 
 A further finding is that slow vibrational motion in the ground state can be considered frictionless. 
\end{abstract}
\maketitle

\section{\label{sec:introduction}Introduction}
Helium nanodroplet isolation (HENDI) spectroscopy allows to study 
atoms, molecules and clusters 
embedded in an ideal cryogenic environment \cite{Toenn_2622_2004}. 
Ultracold helium droplets ($0.38\mathrm{\,K}$) provide
a gentle, since weakly disturbing host for embedded species, 
which can be studied with high resolution spectroscopy \cite{Stien_127_2006}. 
However, spectra of immersed species are slightly broadened and shifted 
away from their gas phase value due to their interaction with the 
surrounding He droplet. 
Inhomogeneous broadening in pure rotational spectra is 
ascribed to the motion of the purity inside the droplet \cite{Calle_4636_2000, Lehma_645_1999} or to coupling to collective degrees of freedom of the droplet \cite{Calle_10090_2001}. 
The overall spectral features, which 
are unseen in gas phase measurements, have been used to study the weak interaction between 
dopant and helium droplet \cite{Reho_161_1997,Grebe_2083_1998, Calle_5058_1999,Calle_4636_2000, Nauta_45084514_2001, Dick_10206_2001}. 
Attached species allow to probe the peculiar properties of the superfluid 
He droplet itself \cite{Hartm_4560_1996}. 
Further insight into the interaction and the quantum properties 
of the droplet is gained from recently obtained 
time-resolved studies \cite{Claas_1151_2006,Schle_245_2010,Mudri_42512_2009, Grune_6816_2011}.

Helium nanodroplets, typically consisting of
several thousand $^4\mathrm{He}$ atoms,  
are ideally suited to study relaxation (cooling) 
of embedded species \cite{Koch_35302_2008, Braun_253401_2004, Przys_21202_2008, Korni_1437_2010, Dropp_233402_2004}. 
  Whether dissipation plays a role depends on the involved energy scales, 
  the coupling strengths, or on typical time scales in the system and ``bath'' \cite{Nitza_200_1975, Ewing_4662_1987}.
  The group of Miller 
  has studied relaxation of the HF system 
  inside He droplets \cite{Nauta_9466_2000} . 
  They report on 
  ineffective vibrational relaxation due to a mismatch in energy scales. 
  However, rotational relaxation 
  in the immersed HF system is fast and appears 
  as Lorentzian line broadening in the rotational transition spectrum.
  (see also \cite{Calle_10090_2001} and references therein). 
  
  Alkali metal atoms and molecules are known to reside in bubble-like 
  structures on the surface of He droplets \cite{Dalfo_61_1994,Ancil_16125_1995,Stien_3592_1995}. 
  Attached Rb$_2$ dimers in the triplet state reveal the presence of vibrational 
  relaxation on the measurement timescale \cite{Higgi_4952_1998,Grune_6816_2011}. For Lithium dimers interacting with a He environment, 
  vibrational relaxation has been investigated by means of Monte Carlo calculations \cite{Bovin_224312_2008}, 
  including a few He atoms. 
  The corresponding relaxation 
  rates depend on whether the dimer is orientated in-plane or out-of-plane
  with respect to the He surface \cite{Bovin_224903_2009}.

  Femtosecond pump-probe techniques are established tools to analyze the 
  ultrafast vibrational motion in molecular systems 
  \cite{Zewai_12427_1993, Grueb_459_1990, Baume_639_1992}. 
  A first laser pulse excites a coherent wave packet, which is allowed 
  to freely evolve on the respective energy surface. The WP
  is probed by a time-delayed pulse. Since 
  the early studies by the Zewail group \cite{Bowma_297_1989, Grueb_883_1993}, vibrational wave packets
  in various diatomic systems have been studied, such as $\mathrm{\,Na}_2$ \cite{Baume_8103_1991} or 
  $\mathrm{\,K}_2$ \cite{Vivie_7789_1996, Rutz_9_1997, Nicol_7857_1999}. 
  When molecules are located 
  in a solid rare-gas matrix \cite{Karav_814_2003, Guhr_5353_2004}, a suppression of 
  revival structures
  in the pump-probe signal indicate a loss of vibrational coherence. 
  Meier et.~al.~have thoroughly investigated such decoherence in molecular systems placed 
  in a rare-gas environment \cite{Meier_4_2004}, motivated 
  by seminal experiments in the Zewail group \cite{Liu_18666_1996}. 
  For experiments at room temperature with a significant 
  thermal occupation of rotational states, 
  unavoidable 
  coupling between internal vibrational and rotational 
  degrees of freedom may lead to sufficient 
  decoherence to suppress 
  (fractional) revivals \cite{Schle_12111_2008}.
  Experiments with Rb$_2$ dimers on He droplets 
  reveal an ongoing decay of the pump-probe ion yield \cite{Mudri_42512_2009}. The 
  decay has been ascribed to damping and accompanying decoherence of vibrational wave packets \cite{Grune_6816_2011}. 
  In this work we investigate potassium dimers on He droplets, 
  studied experimentally with the pump-probe technique \cite{Claas_1151_2006}. 
  A brief account of a theoretical description based on dissipative quantum dynamics was 
  given in \cite{Schle_245_2010}. 
  Experimental spectra show significant deviations from corresponding gas phase calculations for the unperturbed dimer. 

  In this paper, 
  in a phenomenological approach, we investigate in detail how the helium influence may be described 
  effectively. 
  First, we see that the helium environment destroys electronic 
  coherence, which is imprinted by the exciting laser pulse. 
  As possible causes for electronic decoherence, we consider a distribution of shifts of electronic surfaces. 
  Electronic decoherence alone cannot account for the decay of the signal at certain excitation wavelengths. 
  Therefore, in a next step, we include a general damping of vibrational wave packets on each electronic surface. 
  The effective dynamics is described 
  by means of a quantum optical master equation.

  Moreover, it is important to take into account the desorption of dimers from the droplet properly. 
  No general rule can be given when attached atoms or 
  molecules leave the droplet. 
  It is known that lighter alkali metal atoms leave the droplet upon electronic excitation \cite{Stien_10119_2001} or 
  form a bound exciplex \cite{Reho_161_1997a}. The exciplex tends to desorb off the droplet surface during the formation process or several 
  picosecond thereafter \cite{Schul_153401_2001}.
  On the other hand, desorption of Rb atoms may be completely inhibited upon electronical
  excitation in a certain laser wavelength range \cite{Auboc_35301_2008}. 
  Recent measurements indicate that 
  K$_2$ molecules desorb several picoseconds after laser excitation \cite{Claas_1151_2006}. 
  Indeed, (stochastic) desorption of dimers are a crucial ingredient to explain 
  spectral features with a theoretical model \cite{Schle_245_2010}. 
  
  We here report on a more realistic description of desorption - we use 
  a model which only allows electronically excited molecules to leave the droplet. 
  We thus extend our previous, state-independent 
  desorption scheme \cite{Schle_245_2010}. 
  Together with dissipation 
  and shifts of potential energy surfaces, 
  one can explain experimental findings over the full laser excitation range reported in \cite{Claas_1151_2006}. 
  
 The article is organized as follows: 
 In section \ref{sec:pump-probe-signal-1} we 
 review the calculation of the pump-probe signal 
 for free dimers. 
 In the following section \ref{sec:he-influence}, 
 the influence of the helium environment is considered in a phenomenological way. 
 In section \ref{sec:state-indep-desorpt} 
 we explain how we treat desorption of 
 dimers. 
 Finally, for an even better agreement with experiment, 
 we consider undamped motion in the electronic ground state. 
 A thorough comparison between theoretical and experimental 
 findings is given at every step in the model.  
 Section~\ref{sec:conclusion} is devoted to the conclusions. 

\section{\label{sec:pump-probe-signal-1}Pump-probe signal}
\begin{figure}[t]
  \begin{center}    
    {
      \includegraphics[width=.49\textwidth]{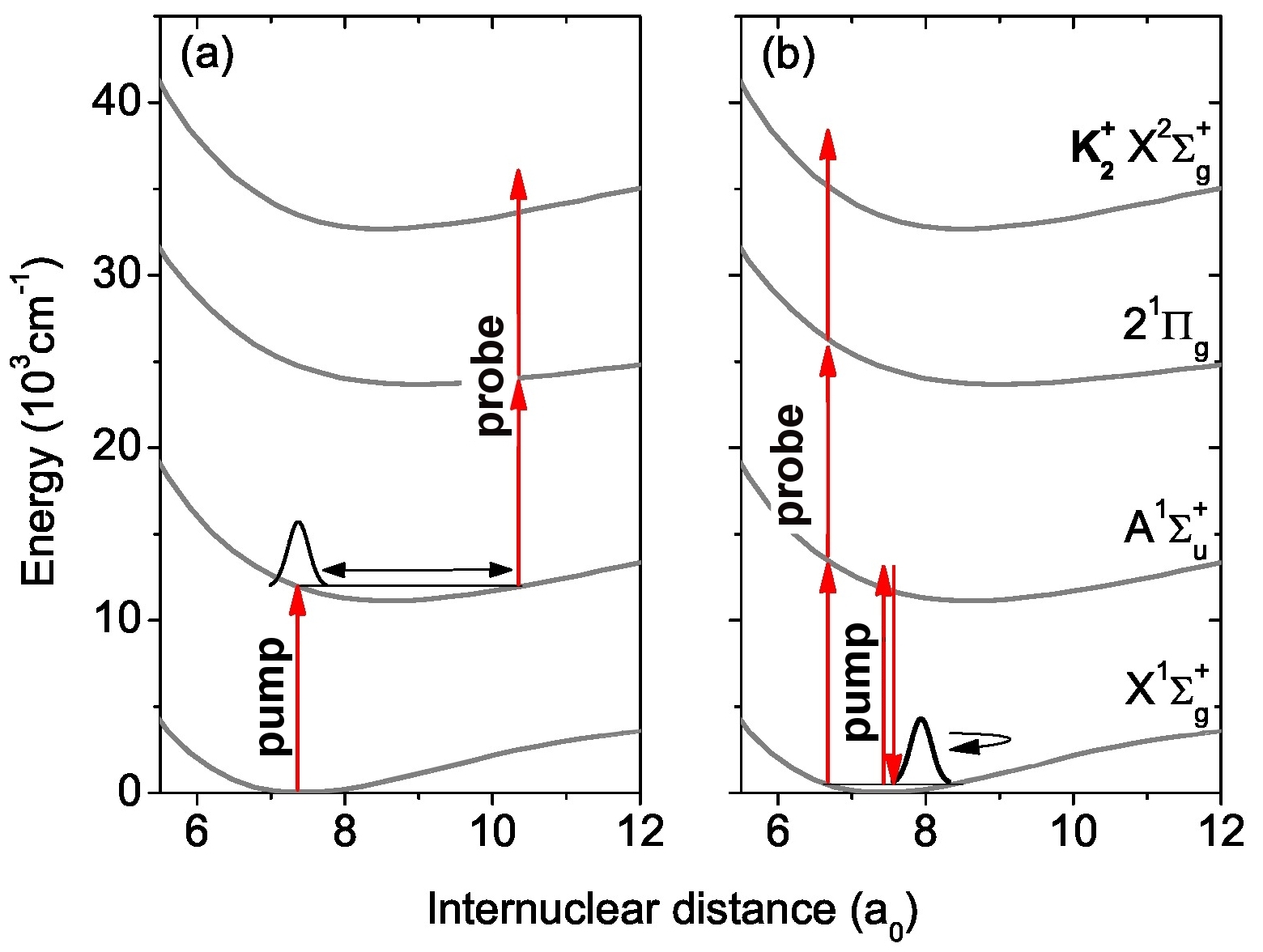}
      \caption{\label{fig:exitationscheme} 
        Excitation schemes in the potassium dimer for two distinct 
        laser wavelengths $\lambda=833\mathrm{\, nm}$ (scheme I) and $\lambda=800\mathrm{\, nm}$ (scheme II). 
        While the first scheme exclusively maps the WP in the $\mathrm{\,A}\, ^1\Sigma^+_u$ state, 
        the second scheme allows to observe the WP in the ground state $\mathrm{\,X}\, ^1\Sigma^+_g$. }}
    \end{center}
\end{figure}

A first ``pump'' laser pulse creates a vibrational wave 
packet (WP) $| \psi_i \rangle$ in some electronic state $|i\rangle$. 
The excited WP oscillates in a region between classical inner and 
outer turning point of that surface. 
It may periodically enter and leave a transition region, where 
 a resonance condition with higher lying states is met 
 and the potential energy difference matches the second, ``probe'' pulse energy. This region 
 defines the so-called Franck-Condon (FC) window. 
 The time-delayed probe pulse leads to a significant 
 number of ions only when the WP is located in the FC window. 
 By varying the time delay $\tau$ between pump-and probe pulse, one obtains an 
 oscillatory ion yield.  
 A typical excitation scheme is depicted in fig.~\ref{fig:exitationscheme}(a), 
where one probes the dynamics of the WP on the $\mathrm{\,A}\, ^1\Sigma^+_u$ surface.

 Wave packets may be excited in several involved electronic 
 surfaces of the dimer. 
 More specifically, the pump pulse prepares the dimer in a superposition 
 of electronic states, such that the full state vector $| \Psi\rangle$
 takes the form 
 \begin{equation}\label{eq:16}
   | \Psi(t_{\mathrm{pump}}) \rangle =\sum_i |\psi_i \rangle |i \rangle .  
\end{equation}
Here, 
 $\psi_i$ denotes the WP on a specific electronic state $i$ and 
 \begin{equation}\label{eq:20}
   p_i=  \langle \psi_i | \psi_i \rangle
 \end{equation}
 is the probability that the electronic state $i$ is excited. 
 
 For the pump-probe signal, we 
 fully numerically solve the time-dependent Schr\"odinger equation (TDSE) 
 
\begin{eqnarray}\label{eq:9}
i \hbar \frac{\partial} {\partial t} \left (     
\begin{array}{ccccc }  \psi_0 \\ \psi_1 \\ \psi_2 \\ \psi_3(E)
\end{array}
\right) \!\! 
 =  \!\! \left( \!\! 
    \begin{array}{ccccc }  
     H_0  &  J_{01} & 0 & 0  \\
      J_{10} & H_1 & J_{12} & 0 \\
      
      0      &  J_{21} &  H_2 & J_{23}^{I} \\
      0      &       0&    J_{32}^{I}  & H_{3,E}
    \end{array} \!\! \right) \!\!   \left (     
\begin{array}{ccccc }  \psi_0 \\ \psi_1 \\ \psi_2 \\ \psi_3(E)
\end{array}
\right), 
\end{eqnarray}
for the full state vector $| \Psi\rangle$ (see also 
\cite{Vivie_16829_1995, Vivie_7789_1996}).
In the matrix equation, the diagonal elements denote the molecular Hamiltonian 
$H_\mathrm{mol}=\sum_i H_i=  T + \sum_i V_{i}$, which involves 
 kinetic energy $T = P^2/2\mu$ (reduced mass $\mu$) 
and adiabatic potentials $V_{i}$. 
In the final ionic state, the 
energy $E$ of the ejected electron is included 
in the diagonal entry of the Hamiltonian $H_{3,E}$, such that $V_{i=final}=V_3+E$. 
The coupling to the laser field 
is described by the matrix elements 
$J_{ij}=-\vec{\mu}_{ij}\cdot \vec{E}(t)$, where $\vec{\mu}_{ij}$ denotes the 
transition dipole moment. 
Both pump- and probe pulse have the form $\vec{E}_\mathrm{pump/probe}(t) = 
  \vec{\epsilon}_0\varepsilon(t)\cos{(\omega_L t)}$ and $\omega_L$ is the 
respective laser frequency. Moreover, 
$\vec{\epsilon}_0$ is the polarization and 
$\varepsilon(t)$ the shape function of the field, 
which is assumed to be Gaussian. 
For the field 
parameters we use a full width at half maximum of 
the laser pulse of $110 \mathrm{\, fs}$ and an intensity of $1.2 \mathrm{\, GW}/\mathrm{cm}^2$. 
The employed intensity is higher than the experimentally estimated value (I=$0.5 \mathrm{\, GW}/\mathrm{cm}^2$ \cite{Claas_1151_2006}), but, according to \cite{Vivie_7789_1996}, still located in the moderate power regime.
The ionic state $\psi_{3}(E)$ also depends on the
energy $E$ of the ejected electron. 
We use discretization of the (electronic) continuum, a technique 
successfully employed earlier \cite{Vivie_16829_1995}. 
 One determines the final state probability $|\psi_{3} (E_k)|^2$ 
after both pulses have passed for distinct electronic energies $E_k$.
The pump-probe signal is 
proportional to the sum over different electronic contributions, 
\begin{equation}\label{eq:13}
S(\tau) = \lim_{t\rightarrow \infty} \sum_{E_k} |\psi_3 (\tau,E_k)|^2.
\end{equation}
In the limit $t \rightarrow \infty$ (upon complete decay of the second pulse), it 
only depends on the delay $\tau$ between the pulses. 
For the propagation of the wave function, we use 
the split-operator method \cite{Feit_412_1982}. 

The ion signal is composed 
of beat frequencies $\omega_{v v'} \equiv (E_v - E_{v'})/ \hbar$ between all pairs 
of energy levels that contribute to the WP \cite{Grune_6816_2011}. 
The most prominent oscillation originates 
from the energy spacing between central and neighboring vibrational levels $\bar{v}$ and $\bar{v} \pm 1$. 
This oscillation has the frequency 
\begin{equation}\label{eq:12}
  \omega_i \equiv  \omega_{\bar{v},\bar{v}+1}
\end{equation}
and is characteristic for the electronic surface $i$. 
In an anharmonic potential, the level spacing and 
therefore also $\omega_i$ decreases as $\bar{v}$ increases. 
 Higher-order frequency components 
$\omega_{\bar{v},\bar{v}+\Delta v}$ with $\Delta v > 1$ are visible in the Fourier transform (FT) of the  signal. 

The laser wavelength $\lambda$ determines which wave packets $\psi_i$ 
can be mapped to the final state. 
In the one-color pump-probe setup, we consider
two different excitation schemes. 

For $820\mathrm{\, nm} \lesssim \lambda \lesssim 840\mathrm{\, nm}$ (scheme I), 
one exclusively follows the vibrational dynamics in the excited state $\mathrm{\,A}\, ^1\Sigma^+_u$. 
The WP in that state can be probed at the outer turning point 
(see fig.~\ref{fig:exitationscheme}(a)). Contributions from 
other surfaces are negligible. For this scheme 
we concentrate on an excitation wavelength  $\lambda=833\mathrm{\, nm}$.      
\begin{figure}[t]
  \begin{center}
    {
      \includegraphics[width=.49\textwidth]{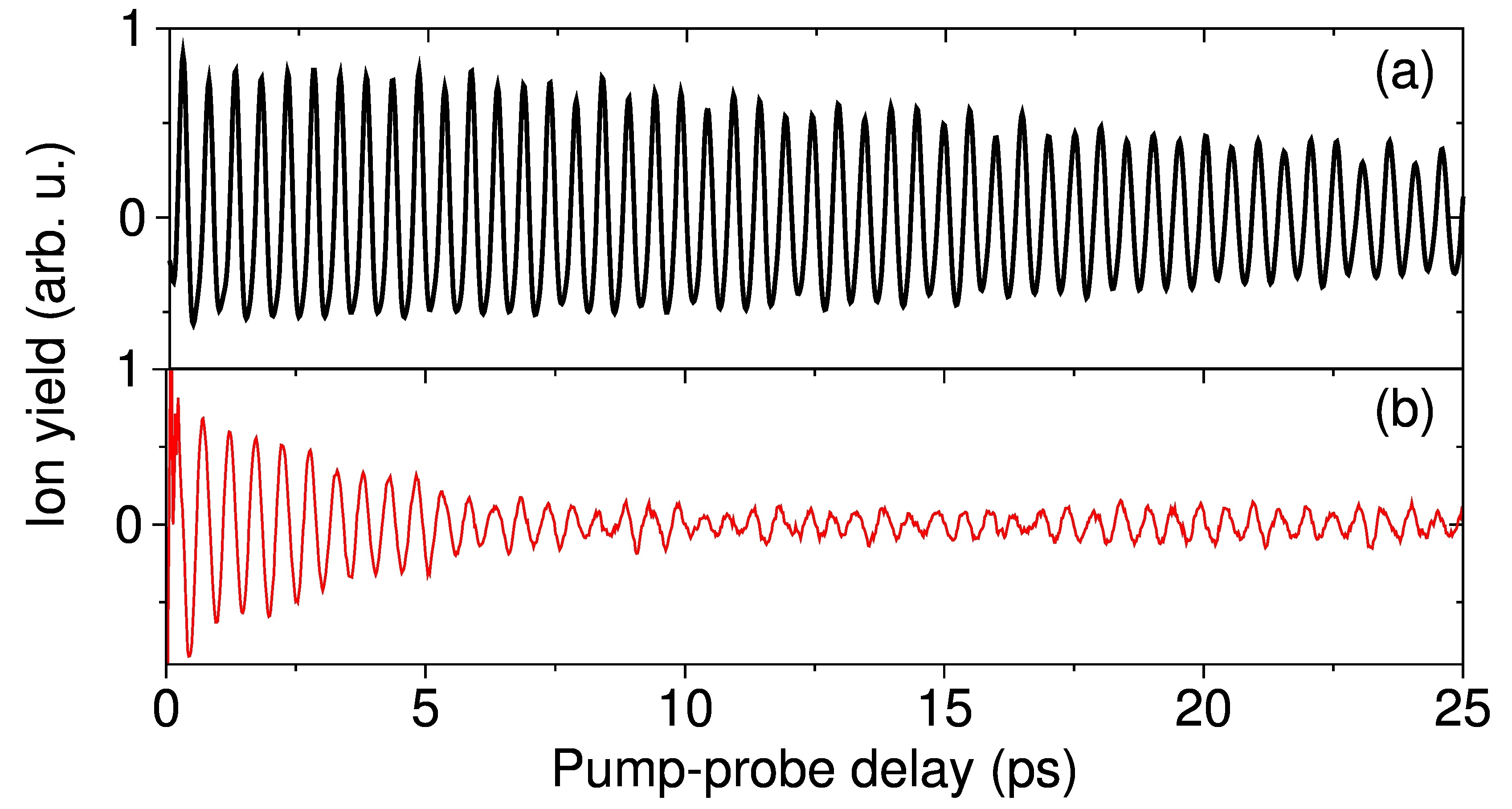}
      \caption{\label{fig:PP_833_gphase_exp}Pump-probe ion yield at $\lambda=833\mathrm{\, nm}$. (a) 
        Numerical gas phase calculation. (b) Experimental result (from \cite{Claas_1151_2006}). 
        The oscillation is exclusively attributed to the 
        circulating WP in the $\mathrm{\,A}\, ^1\Sigma^+_u$ state. 
      }}
  \end{center}
\end{figure}
Fig.~\ref{fig:PP_833_gphase_exp}(a) shows the theoretical 
gas phase calculation. Fig.~\ref{fig:PP_833_gphase_exp}(b) shows experimental spectra at this wavelength 
 obtained from dimers attached to helium nanodroplets \cite{Claas_1151_2006}.  
 In the experiment, the signal amplitude significantly decreases, but 
 oscillates with nearly constant amplitude at later delay times.

 \begin{figure}[b]
   \begin{center}
     {
       \includegraphics[width=.49\textwidth]{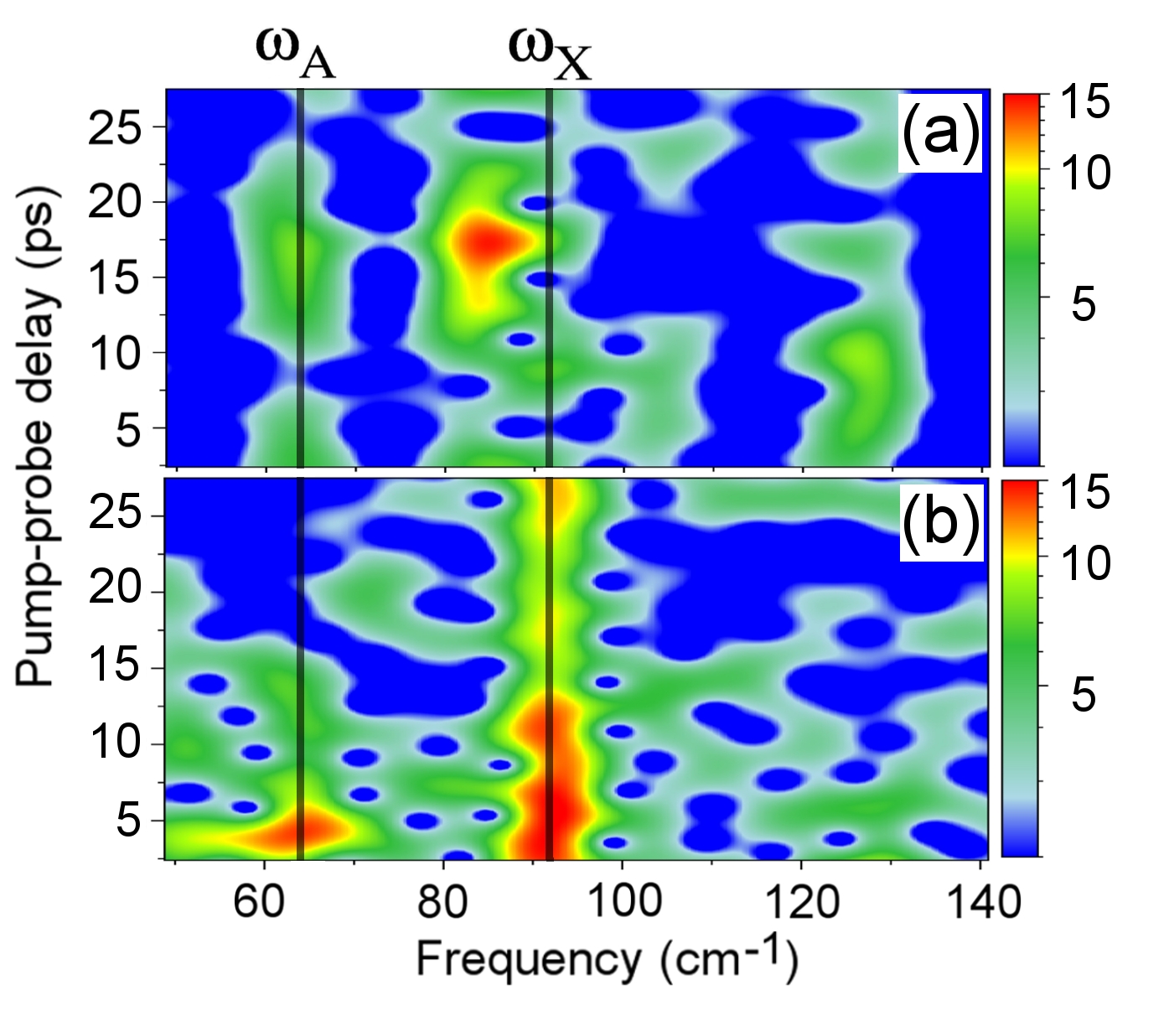}
       \caption{\label{fig:FT_800nm_gphase_exp}
         Spectra $\mathcal{F}(\omega,\tau)$ in the 
         time-frequency domain at $\lambda=800\mathrm{\, nm}$. 
         (a) Numerical gas phase calculation, where electronic interferences 
         lead to a spectrum that is difficult to interpret. 
         (b) Experimental HENDI result (from \cite{Claas_1151_2006}). 
         Electronic interferences are absent, allowing for a clear identification of the structures of the spectrum. 
       }}
   \end{center}
 \end{figure}
 For $800 \lesssim \lambda \lesssim 820\mathrm{\, nm}$ (scheme II), transitions 
 preferably take place at the inner turning point of the excited WP. 
 The wave packet in the electronic 
 ground state $\mathrm{\,X}\, ^1\Sigma^+_g$ is excited 
 through resonant impulsive stimulated Raman scattering (RISRS). It can be probed through a 3-photon process, see fig.~\ref{fig:exitationscheme}. 
 On the other hand, simultaneous and coherent contributions 
 from WPs in electronic excited states lead to constructive or 
 destructive interferences and thus to an unstructured ion yield \cite{Nicol_7857_1999}. 
 We use sliding window Fourier transforms $\mathcal{F}(\omega, \tau$) 
 to follow the evolution of respective 
 beat frequency components as a function of the delay time \cite{Fisch_331_1996, Vrakk_37_1996}.
 Fig.~\ref{fig:FT_800nm_gphase_exp}(a) shows a window transform (spectrogram) for excitation at $\lambda=800\mathrm{\, nm}$. A
 frequency component $\omega_A \approx 63 \mathrm{\,cm}^{-1}$ can be ascribed to the 
 WP in the state $\mathrm{\,A}\, ^1\Sigma^+_u$.  Moreover, a higher-order frequency component around $\approx 2\omega_A$ is visible as  is a significant contribution at $\omega \approx 85\mathrm{\, cm}^{-1}$ after about 15 ps that can be traced back to originate from the $2 ^1\Pi_g$ surface. 
 A contribution 
 from the ground state and resulting frequency component $\omega_X$, however, is missing. 
In agreement with earlier 
 findings \cite{Nicol_7857_1999, Vivie_7789_1996}, the mapping of the ground state WP is not possible at low to intermediate intensities in the gas phase. 
 This is because, as explained below, contributions from different potential energy surfaces interfere destructively. 
 In contrast, for dimers attached to He droplets, 
 the contribution from the state $\mathrm{\,A}\, ^1\Sigma^+_u$ is suppressed after about 8 ps, while 
 on the other hand the vibrational ground state WP is clearly resolved, 
 see fig.~\ref{fig:FT_800nm_gphase_exp}(b).

\begin{figure}[t]
\begin{center}
{
\includegraphics[width=.49\textwidth]{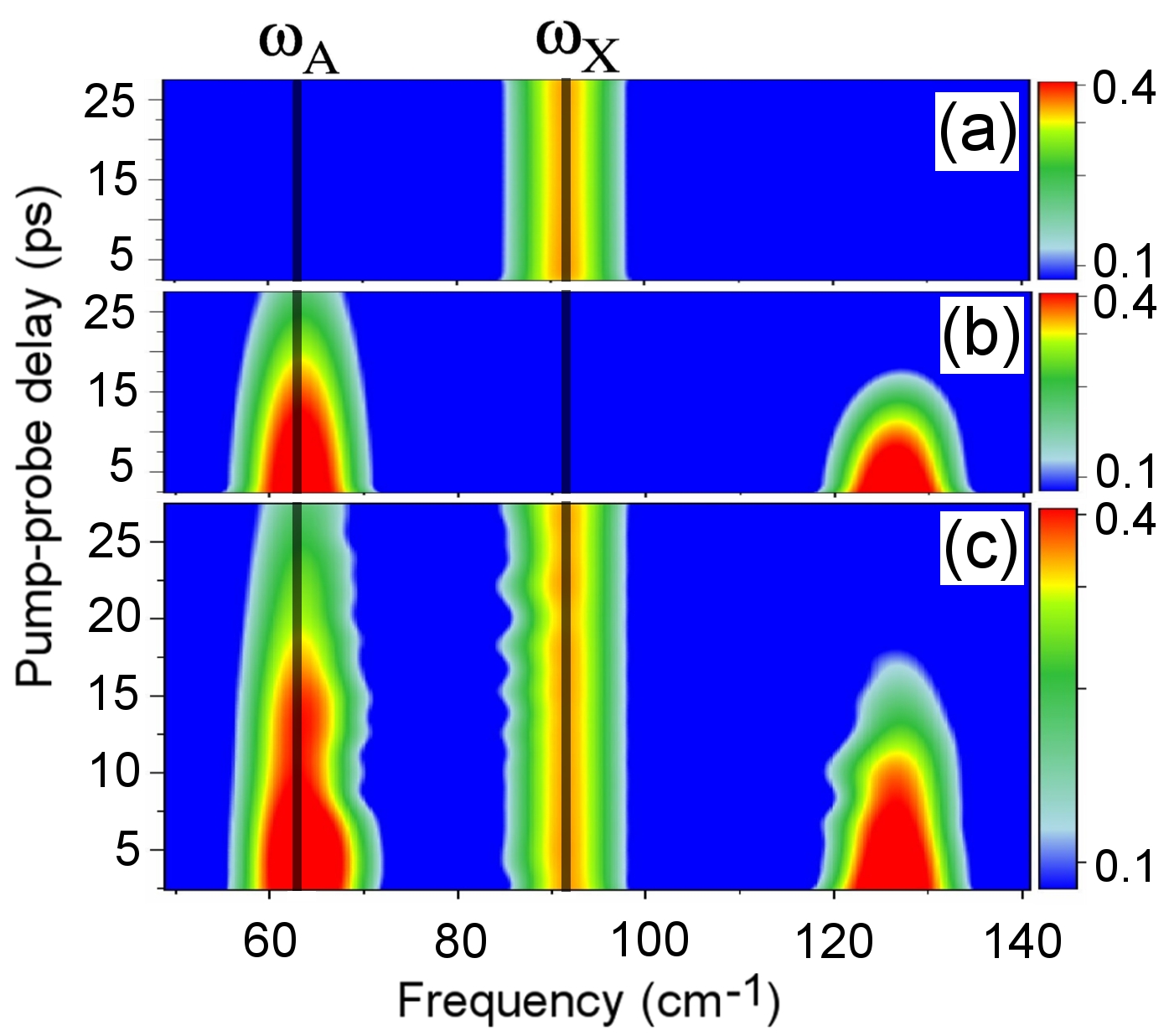}}
\caption{\label{fig:proj_scheme}
  Shown is a the spectrogram $\mathcal{F}^i(\omega,\tau$), which is obtained 
  upon projecting the created full wave function $\Psi(t)$ on the electronic ground 
  state ($i=0$) and (b) first excited state ($i=1$). 
  (c) In the incoherent sum, interferences are excluded.  
  In particular, the ground state component is clearly resolved.   
}
\end{center}
\end{figure}

 Qualitatively, the difference between fig.~\ref{fig:FT_800nm_gphase_exp}(a) (gas phase) and fig.~\ref{fig:FT_800nm_gphase_exp}(b) 
 (experiment) can be 
 explained by the loss of electronic coherence alone:
 In fig.~\ref{fig:proj_scheme}, we construct an artificial signal from  
 the incoherent sum of contributions. 
 The full coherent wave vector after 
   decay of the pump pulse at $t\equiv t_\mathrm{pump}$, eq.~(\ref{eq:16}), serves 
 as a starting point. 
 In order to determine the contribution of a single WP $\psi_i$, we 
 project the fully coherent wave vector according to  $\tilde{\Psi}_i(t_\mathrm{pump})= P_i| \Psi(t_\mathrm{pump})\rangle $ with the projector $P_i=|i \rangle \langle i|$ on one of the electronic states.  
 The electronic 
 occupation $p_i$ is not altered and 
 the usual probe scheme is employed after projection.
 The resulting spectrogram $\mathcal{F}^i(\omega, \tau)$ 
 after projection on the state $\mathrm{\,A}\, ^1\Sigma^+_u$ and $\mathrm{\,X}\, ^1\Sigma^+_g$, respectively, is shown 
 in fig.~\ref{fig:proj_scheme}(a)/(b). The incoherent sum of 
 contributions is given by 
 \begin{equation}\label{eq:17}
   \mathcal{F}^{\mathrm{inc}}(\omega,\tau)=\sum_i \mathcal{F}^i(\omega, \tau). 
 \end{equation}
 Contrary to the coherent signal, fig.~\ref{fig:FT_800nm_gphase_exp}(a), 
 in eq.~(\ref{eq:17}) interferences are removed by hand and contributions from ground- and 
 excited state WPs are clearly visible. Obviously, the incoherent sum fig.~\ref{fig:proj_scheme}(c) already 
 reproduces important features of the experimental data. 
In the experiment (fig.~\ref{fig:FT_800nm_gphase_exp}(b)), however, the component $\omega_A$ fades out at later 
delay times, while the component $\omega_X$ becomes dominant. 
Therefore, a model based on pure loss of electronic coherence alone 
cannot explain the measured spectrum. 
One has to consider additional influences of the He environment 
on the dimer dynamics.

\section{He influence and results}\label{sec:he-influence}

 In a phenomenological approach, 
 we take into account three possible 
 effects:
 \begin{enumerate}
   \item A He-induced energetic shift
     of electronic potential energy surfaces, with possibly small fluctuations. 

   \item Damping of vibrational wave packets. 
     This effect is treated fully quantum 
     mechanically within the master equation 
     approach. The WP $\psi_i$ dissipates energy with a 
     certain damping rate $\gamma_i$. This rate is here seen as a fit parameter and is  
     adjusted to experimental observations.

     \item   Desorption of dimers off the droplet. 
       After desorption, the influence of the helium droplet 
       (shift/damping) vanishes. 

  \end{enumerate}
  A description based on damping 
  of vibrational wave packets has been 
  applied to HENDI studies 
  with spin triplet Rb$_2$ dimers \cite{Grune_6816_2011}.
  There, slow vibrational decoherence (as a consequence of very weak dissipation) 
  is most relevant for the decay of the revival amplitude. 
  In particular, as witnessed by the ongoing decay, desorption of dimers seems to be inhibited. 

  In the following, step by step, we include shift, damping 
  and finally desorption 
  in the calculation of pump-probe spectra. 
  It is found that all model ``ingredients'' 1.~-~3.~are crucial 
  to find agreement with obtained experimental spectra.

\subsection{Electronic shifts}\label{sec:electronic-shifts}
 The He environment may lead to shifts of electronic surfaces in attached dimers. 
 Such shifts are common for alkali atoms and molecules on He droplets \cite{Higgi_4952_1998}.
 Spectra from attached species 
 are shifted relative to what one expects from gas-phase potential energy curves \cite{Stien_253_1996, Allar_1169_2006, Bruhl_10275_2001, Auboc_54304_2010, Buener_12684_2007}. For certain weakly coupled species, 
 spectra are only shifted by a few wavenumbers \cite{Higgi_4952_1998}.
 Denoting the shift of surface $i$ with $\Delta_i$, the full state vector is propagated according to
\begin{equation}\label{eq:7}
    |\dot{\Psi}(t) \rangle = -\frac{i}{ \hbar} H |\Psi(t) \rangle + \frac{i}{ \hbar} \sum_i \Delta_i  | \psi_i \rangle
\end{equation}
Both surfaces $\mathrm{\,A}\, ^1\Sigma^+_u$ and 2$^1 \Pi_g$ may be affected. 
Only relative shifts are relevant for the signal, 
such that we cannot differentiate between a shift of the electronic ground - or first excited state.
Also, a shift of the ionic surface would change the 
energy distribution of the ejected electrons, but not the overall calculated ion yield. 
This is because we sum over all electronic energies in eq.~(\ref{eq:13}). 
We find that  even large, but fixed shifts of up to $\pm  100 \mathrm{\, cm}^{-1}$ 
have a very small influence and resemble the gas phase calculation. 

However, it is reasonable to assume that the shifts $\Delta_i$ fluctuate for the following reason:
The number of He atoms of a droplet is not fixed but varies according to a log-normal distribution \cite{Lewer_381_1993}. 
Consequently, we have to average over a distribution of shifts $\Delta_i$ \cite{Login__2008}. 
Also, for the considered laser intensities, 
electronic surfaces are slightly Stark-shifted. 
Depending on their position in the laser beam, dimers are exposed to a distribution of 
laser intensities \cite{Mudri_priv}. The beam 
 width leads to a distribution of Stark shifts, 
 which in turn has to be treated as random distribution of electronic shifts $\Delta_i$. 

Considering fluctuating potential energies, eq.~(\ref{eq:7}) represents only 
a single realization $|\Psi_j(t)\rangle$ with shifts $\Delta_i(j)$ in the ensemble.
 The pump-probe signal is proportional to the ensemble average 
\begin{equation}\label{eq:14}
 \langle S(\tau) \rangle = \frac{1}{N} \sum_j \left [\lim_{t\rightarrow \infty} \sum_{E_k} |\psi_{j,3} (\tau,E_k)|^2\right ], 
\end{equation}
where $N$ realizations have been taken into account. 
In this expression, the final state $ |\psi_{j,3} \rangle$ is obtained from propagation of the realization  $|\Psi_j(t)\rangle$, 
in which  $\Delta_i(j)$ is chosen randomly.
Through the ensemble average in eq.~(\ref{eq:14}), one 
obtains an incoherent mixture of electronic contributions in the final state. 
The spectrogram of $\langle S(\tau)\rangle$ nearly perfectly resembles the incoherent 
sum of contributions, see eq.~(\ref{eq:17}) and fig.~\ref{fig:proj_scheme}(c). 
 It does not contain any electronic coherence, such that the frequency components $\omega_i$ are clearly resolved. 
 For full decoherence fig.~\ref{fig:proj_scheme}(c), 
 random shifts in the range of $\pm 5\mathrm{\, cm}^{-1}$ around 
 the average shifts $\overline{\Delta_1}=0\mathrm{\, cm}^{-1}$ and $\overline{\Delta_2}=-50\mathrm{\, cm}^{-1}$ are sufficient. 
 Note that this decoherence is due to the inhomogeneous size distribution and not 
 due to entanglement between system and environment \cite{Schlo__2007, Helm_42108_2009}.
  
 To conclude, randomly distributed 
 energy shifts can explain the 
 visibility of the ground state component $\omega_X$. 
 However, this effect lacks an 
 explanation for the decay of the frequency component $\omega_A$ of the 
 excited state, which is observed in the HENDI experiment. Therefore, we consider vibrational damping next. 

\subsection{Damped vibrational wave packets}\label{sec:damp-vibr-wave}

 Dissipation can be treated fully quantum mechanically 
 by using approximate 
 master equations for the density operator
 of the (reduced) system. 
 The overall dissipation of an excited WP 
 originates in our case from the interaction of 
 vibrational degrees of freedom of the molecule 
 with collective degrees of freedom of the helium droplet. 
 Note that in recent studies \cite{Bovin_224903_2009} vibrational relaxation rates for alkali dimers 
 on $^4 \mathrm{He}$ clusters are estimated from full quantum Monte Carlo calculations for 
 a few He atoms. 
 These rates turn out to be roughly of the same order of magnitude as our phenomenologically chosen 
 damping rates below. 

 In our approach, the density $\rho$ of the damped WP in an 
 electronic state $|i \rangle$ 
 evolves according to the master equation
 \begin{equation}\label{eq:3}
   \dot{\rho} = - \frac{i}{\hbar} [H_i,\rho] 
   +\gamma \left( a \rho a^{\dagger}  
   -\frac{1}{2} a^\dagger a \rho - \frac{1}{2} \rho a^\dagger a \right ), 
\end{equation}
which is of Lindblad form \cite{Lindb_147_1975}. It describes 
friction for near-harmonic systems at effectively zero temperature in the 
rotating wave approximation (see, for instance, \cite{Scull__1997}). 
The first term in eq.\,~(\ref{eq:3}) contains the molecular Hamiltonian $H_i$
of a specific electronic state $|i\rangle$ and determines the unitary evolution of the WP $| \psi_i \rangle$. 
 Relaxation of the vibrational WP is 
achieved through the second, irreversible contribution. Any initial 
state approaches the ground state on a time scale $\gamma^{-1}$, 
i.\,e.\,$\gamma$ denotes the damping rate. 
In eq.~(\ref{eq:3}), $a, a^\dagger$ are the 
creation/annihilation operators of a harmonic oscillator, defined through
\begin{equation}\label{eq:10}
a = \frac{1}{\sqrt{2}}\left ( \sqrt{\frac{\hbar}{m\omega_e }}\hat{X} 
+ i\frac{1}{\sqrt{\hbar m \omega_e }} \hat{P} \right). 
\end{equation}
Here, $\hat{X}$ and $\hat{P}$ are 
the usual position and momentum operator 
w.r.t. the harmonic oscillator minimum and $\omega_e$ its frequency. 
 The quantum optical master equation (\ref{eq:3}) can be derived from a von-Neumann equation for the full system and is valid only for weak couplings between system and ``bath'' (which is the helium droplet here). 
Also, in the derivation one makes use of the 
Markov and rotating wave approximation. 
For a significant temperature, additional terms that describe thermal excitations from the 
environment have to be taken into account \cite{Breue__2002}. 
The damping constant $\gamma$ may in principle be derived from a 
microscopic description of interaction 
between system and bath (Fermi's Golden Rule). We do not specify this interaction, but use the damping 
rate $\gamma$ as a fit parameter to obtain agreement with experimental data. 
The master equation (\ref{eq:3}) induces the 
evolution of a pure initial state into a state mixture. 

\begin{figure}[t]
\begin{center}
{
  \includegraphics[width=.49\textwidth]{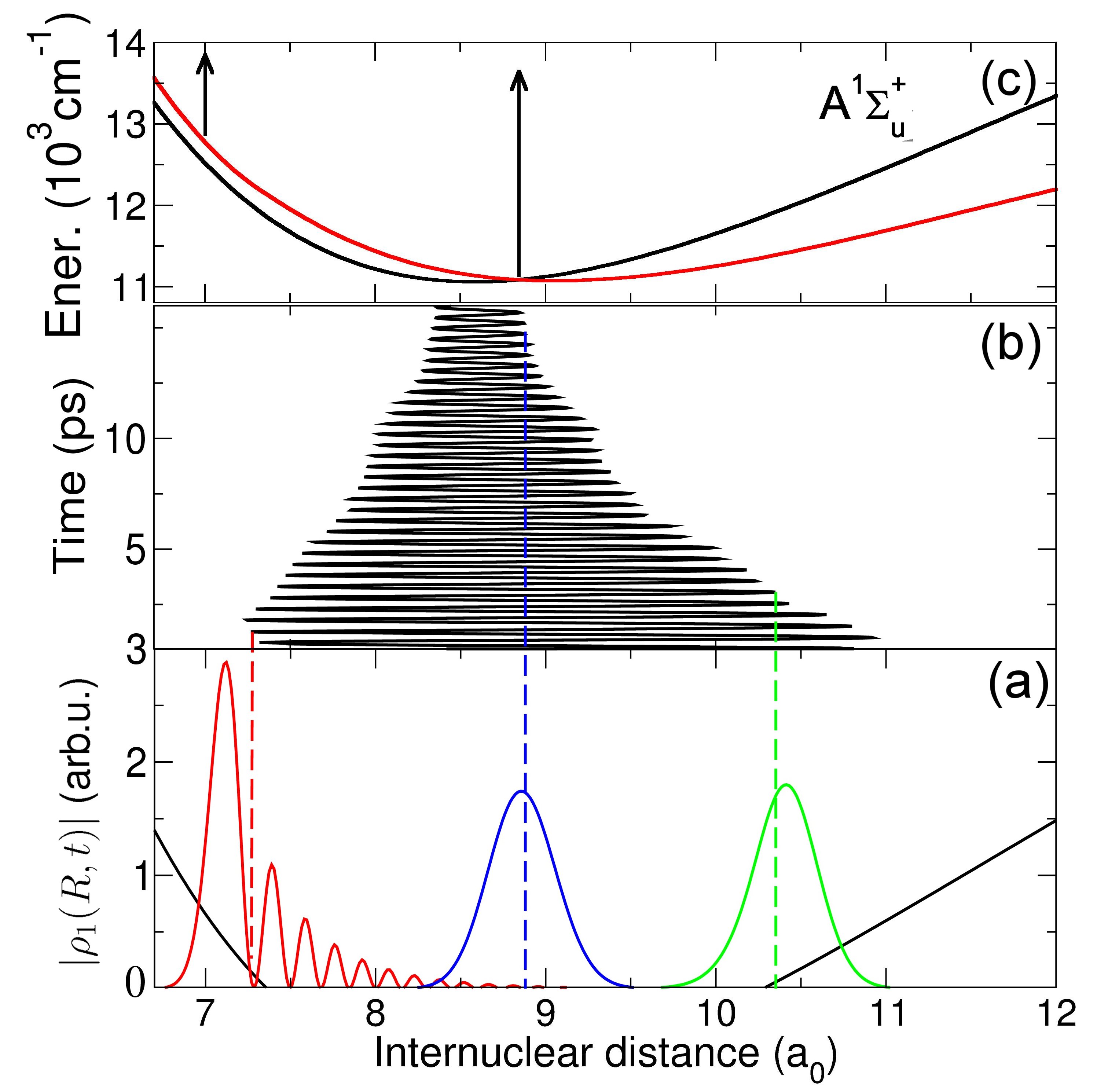}}
\caption{\label{fig:WPs_833nm}
  (a) Shown is single realization of 
  a wave function in position space. The WP circulates 
  in the A state and is damped with a rate $\gamma_1=0.3/\mathrm{ps}$. 
  (b) Coordinate expectation value for that single realization. 
  The noise is due to the stochastic propagation. 
  (c) Potential curves A (black) and 2$\Pi$ (red). 
  The latter is shifted by one 
  photon energy ($\lambda=800\mathrm{\, nm}$) to clarify 
  possible transitions to the ionic state (arrows). 
  A fully damped WP can be mapped to the final state.
}
\end{center}
\end{figure}
For the numerical propagation, we return from the density matrix description 
to a Schr\"odinger-type equation for the state vector. 
It is not possible to evolve a pure state into a mixed state 
with a deterministic Schr\"odinger equation. 
One therefore considers a stochastic differential equation for a state vector, quantum state diffusion (QSD) 
\cite{Gisin_5677_1992}.
 The density is recovered from the average over several realizations of state vectors. 
Given a master equation in ``Lindblad form'', as in our case,
it is straightforward to state the corresponding quantum state diffusion (QSD) Ito stochastic Schr\"odinger equation for a state vector 
\cite{Gisin_5677_1992}: 
\begin{eqnarray}\label{eq:4}
    |d\psi(t) \rangle  = & -\frac{i}{\hbar} H_i |\psi \rangle dt   \\ \nonumber
    & +  \frac{1}{2} \gamma \left(2 \langle a^\dagger \rangle a  
      - a^\dagger a - |\langle a \rangle|^2 \right)
    |\psi \rangle dt  
    + \sqrt{\gamma}(a- \langle a \rangle ) |\psi \rangle d\xi(t). 
\end{eqnarray}
The left hand side means $|d\psi(t) \rangle = |\psi(t+dt) \rangle - 
|\psi(t)\rangle$, i.\,e. the change of the state after a time 
increment $dt$. 
The second term induces transitions to lower lying vibrational states. 
The third term is stochastic and contains complex 
normalized Ito increments $d\xi(t)$, which satisfy 
\begin{eqnarray}\label{eq:5}
   d\xi^2     & =  (d\xi^*)^2  =  0   \nonumber \\
   d\xi d\xi^* & =  dt .
\end{eqnarray}
The increment of the density up to second order in $dt$ is given through 
$ d \rho = | d \psi \rangle \langle \psi| + 
|\psi \rangle \langle d \psi| + | d \psi \rangle \langle d \psi |$. 
Since eq.\,~(\ref{eq:4}) is written in Ito form, 
one can easily take the average with respect to (\ref{eq:5}) and prove the 
equivalence with the given master equation (\ref{eq:3}). 
Recovering the density $\rho=\overline{ | \psi(t) \rangle \langle \psi(t) | }$ 
from several realizations of $|\psi\rangle$ amounts 
to obtaining expectation values through $\langle A \rangle = \mathrm{Tr} ( A \rho ) = \overline{ \langle \psi | A | \psi \rangle} $.
In fig.~\ref{fig:WPs_833nm} we show the coordinate 
expectation value $\langle R \rangle$, 
obtained from a single realization 
of $| \psi \rangle$. 
The norm of the state in eq.~(\ref{eq:4}) is conserved,  i.\,e.\,$d (\langle \psi | \psi \rangle) = 0$. 
However, due to the finite time step $\Delta t$, the norm can slightly fluctuate. 
Therefore, for the numerics, we impose norm preservation by 
renormalizing the state vector after every time step. 
After replacing the ladder operators in eq.\,(\ref{eq:4}) through their definition 
in (\ref{eq:10}), 
the r.\,h.\,s.\, is strictly separable in operators, 
which act in either momentum or coordinate space.
Therefore, the split operator method \cite{Feit_412_1982} 
for the propagation of the WP can still be used, 
which is an advantage of this approach. 

Helium induced damping of vibrational wave packets on 
a {\it single} electronic surface is described through eq.~(\ref{eq:4}), 
 which is equivalent to the master equation eq.~(\ref{eq:3}) on average. 
In order to obtain damping on several electronic 
surfaces, we propagate a full state vector $|\Psi(t)\rangle $ according to 
\begin{equation}\label{eq:15}
  |d\Psi(t) \rangle = -\frac{i}{ \hbar} H |\Psi(t) \rangle dt + \underbrace{\sum_j\left[ D(\gamma_j) + \frac{i }{ \hbar} \Delta_j \right] | \psi_j \rangle}_{\mathrm{\,coupling\,to\,He\,bath}} 
\end{equation}
The average is taken over several realizations of state vectors to 
recover the density via $ \rho(t)= \overline{ | \Psi(t) \rangle \langle \Psi(t) | }$.
In eq.~(\ref{eq:15}), 
$D(\gamma_j)$ is the generalization of the r.h.s. of eq.~(\ref{eq:4}) to obtain 
damping and accompanying fluctuations on an arbitrary surface,  
\begin{equation}\label{eq:1}
D(\gamma_j) = \frac{1}{2} \gamma_j \left(2 \langle a_j^\dagger \rangle a_j  
 - a_j^\dagger a_j - |\langle a_j \rangle|^2 \right) dt 
              + \sqrt{\gamma_j}(a_j- \langle a_j \rangle ) d\xi_j(t) 
\end{equation}

In eq.~(\ref{eq:1}), all stochastic differential Wiener increments $d \xi_j$ are taken independently of 
each other. 
In the ladder operator of surface $j$, one uses the 
the position w.r.t. the respective harmonic oscillator minimum and its frequency. 
For most parts of the following,
for simplicity, we set damping constants to be equal $\gamma_j=\gamma$. 
However, we hasten to add that
an interesting exception is provided in section \ref{sec:undamp-ground-state}. 
The agreement with experiment improves, if one allows for {\it undamped} 
vibrational motion in the electronic ground state, $\gamma_0=0$. 

The damping model provides an explanation for the signal decrease 
at $\lambda=833\mathrm{\, nm}$ (excitation scheme I). 
After several periods, vibrationally damped WPs on the A surface no longer enter the initial FC region 
and therefore the ion yield decreases. 
 \begin{figure}[t]
 \begin{center}
 {
 \includegraphics[width=.49\textwidth]{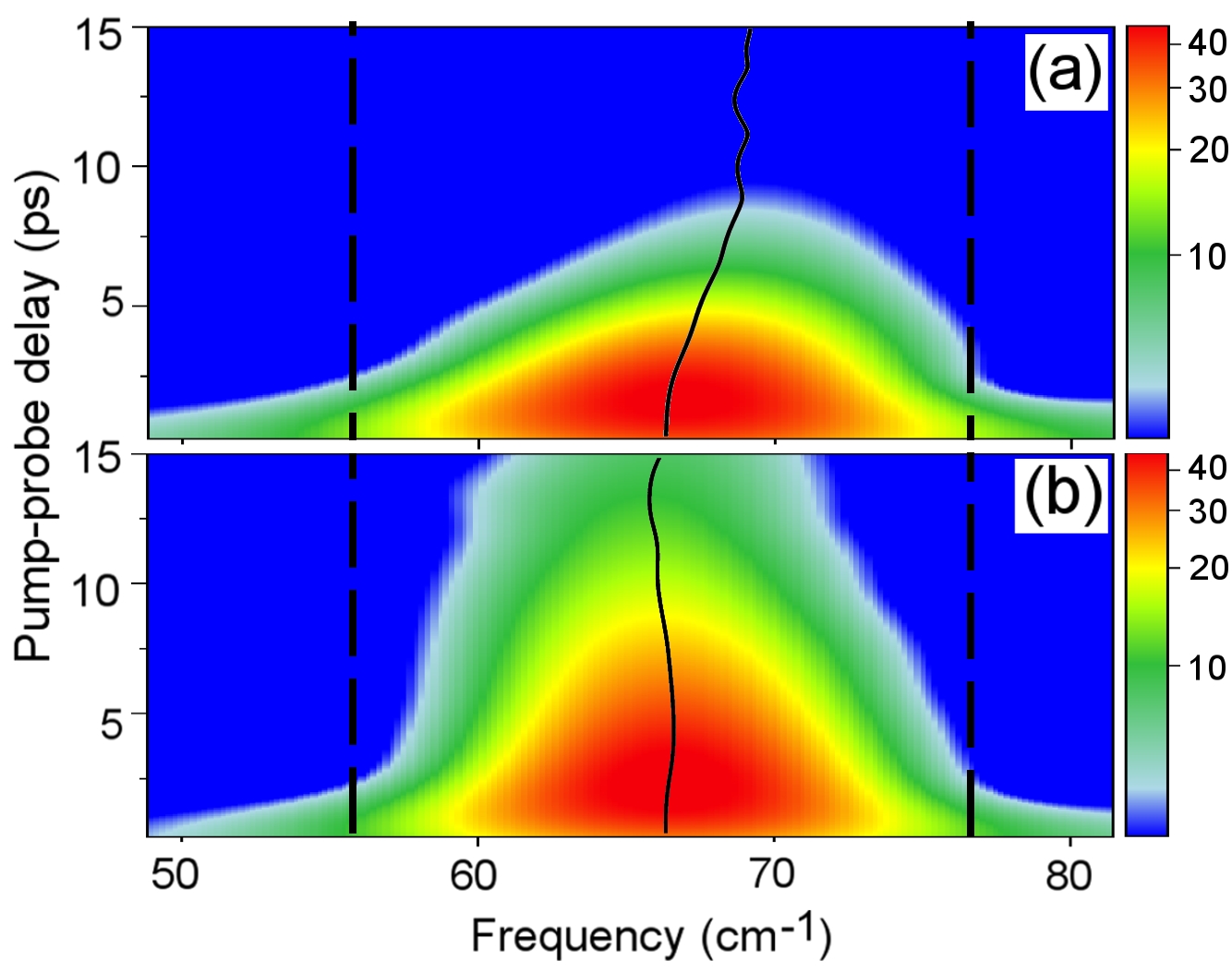}}
 \caption{\label{fig:FT_833nm_SD}
   Spectra $\mathcal{F}(\omega,\tau)$ in the 
   time-frequency domain at $\lambda=833\mathrm{\, nm}$. 
   (a) Full damping. 
   (b) Damping and state-dependent desorption. 
   Also shown is the frequency 
   upon averaging in the lined frequency interval. Explanation see text. 
 }
 \end{center}
 \end{figure}
\begin{figure}[t]
\begin{center}
{
\includegraphics[width=.49\textwidth]{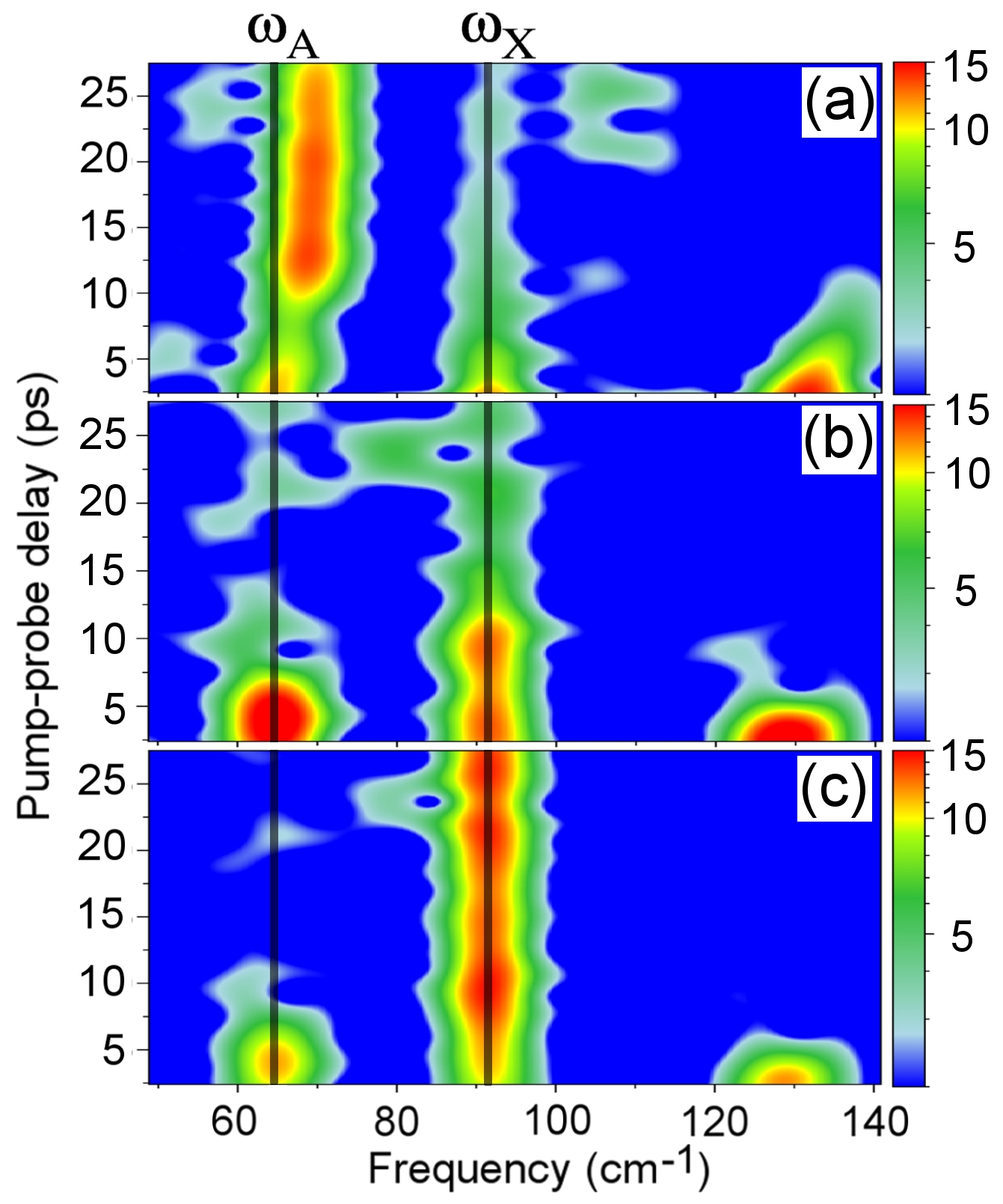}}
\caption{\label{fig:PP_FT_SD_800nm}
   Calculated spectra in the time-frequency domain 
   at $\lambda=800\mathrm{\, nm}$. 
   (a) Full damping: The WP in the A state 
   leaves the initial transition region, consequently the component $\omega_A$ decreases. 
   The fully damped WP, however, reaches a 
   transition region at the equilibrium distance, such that 
   $\omega_A$ returns. See also fig.~\ref{fig:WPs_833nm}. (b) State-dependent desorption model. (c) 
   State-dependent desorption model together with undamped motion in the ground state. 
}
\end{center}
\end{figure}
In fig.~\ref{fig:FT_833nm_SD} we depict the result 
 of the damping model eq.~(\ref{eq:15}), using $\gamma_1=0.15/\mathrm{ps}$. 
 Clearly visible is a 
 shift of the central frequency $\omega_A\rightarrow \omega_A'>\omega_A$, since 
 the WP relaxes to lower vibrational levels.
 A frequency shift is also observed experimentally \cite{Claas_1151_2006}. 
 In the full damping model, however, the signal decays to zero, since at this 
 wavelength vibrational relaxation 
 leads to a complete ``closing'' of the initial FC window. This result 
 is in contrast to the experimental observation, where 
 a pronounced oscillation is also present at later delay times, see \cite{Claas_1151_2006} and fig.~\ref{fig:PP_833_gphase_exp}(b),  and requires further studies below.  

 At $\lambda=800\mathrm{\, nm}$ (excitation scheme II), full damping 
 leads to the result shown in  fig.~\ref{fig:PP_FT_SD_800nm}(a). 
 Through dissipation, the WP in the state $\mathrm{\,A}\, ^1\Sigma^+_u$ leaves the initial FC region. 
 However, the decelerated WP approaches another FC window around the equilibrium distance, cf.~fig.~\ref{fig:WPs_833nm}, 
 after several circulations. 
 The signal therefore 
 shows a massive increase of the (shifted) frequency component $\omega_A'$. 

 For both excitation schemes, the inclusion of damping improves agreement with experiment for the 
 first $\simeq 10\mathrm{\, ps}$. However, at later times, there are still significant 
 discrepancies. These can be removed by taking into account desorption of dimers off the droplet. 
 
 The clear visibility of both frequency components $\omega_A'$ and 
 $\omega_X$ is again attributed to electronic decoherence: 
 Since the damping scheme is carried out 
 (stochastically) independent on all involved 
 electronic states, damping leads to 
 electronic decoherence similar to the fluctuating shifts we assumed previously. 
 The final signal is an incoherent mixture of 
 electronic contributions, as previously considered through eq.~(\ref{eq:17}). 
 However, a massive increase of the component 
 $\omega_A'$ is not observed experimentally (compare fig.~\ref{fig:FT_800nm_gphase_exp}(b) and fig.~\ref{fig:PP_FT_SD_800nm}(a)). 

 Desorption of the dimer off the droplet  
 prevents the full vibrational damping in the electronic ground or excited state. 
 Indeed, as we will show, desorption implies the 
 disappearance of component $\omega_A$ and simultaneous ongoing 
 presence of $\omega_X$. For excitation scheme I, desorption 
 explains the observed signal oscillation at later delay times.

 \subsection{\label{sec:state-indep-desorpt}  Desorption }

 The release of a dimer into the gas phase takes place at some random time $t'$. 
 In our model, a state vector evolves according to eq.~(\ref{eq:15})
 up to time $t'$. 
 As the He influence vanishes at $t'$, shift/damping terms in 
 eq.~(\ref{eq:15}) are set to zero. 
 We have to consider the 
 ion yield at delay time $\tau$, which is changed because of the dissapearance 
 of the He influence at time $t'$. 
 If the desorption occurs before the 
 decay of the  probe pulse ($t' \lesssim \tau$), 
 the resulting signal is denoted as $S(\tau,t')$. 
 If the desorption occurs after decay of the probe pulse ($t' \gtrsim \tau$), 
 the ion yield is unaffected by the desorption process. Anything 
 that happens after the probe pulse, will not be mapped to the ion yield. 
 The resulting signal is denoted with $S(\tau,\tau)$. 
 
 Note that in an ensemble of attached dimers, 
 the desorption time $t'$ will be distributed 
 according to some probability distribution $P(t')$. We therefore 
 have to calculate the pump-probe signal for various desorption times $t'$ 
 and consider the averaged signal. 

We first assume that the electronic state occupation is not relevant in the desorption process. 
A dimer therefore has a certain probability to stay on 
 or leave (still being in the superposition) the droplet,  
 but this probability is independent of the specific electronic occupation. 
 We assume a constant probability 
 \begin{equation}\label{eq:6}
   p_\mathrm{off}(t', t' + \Delta t')=\Delta t' R_D
 \end{equation}
 that the dimer leaves the droplet within a small 
 time interval $(t', t' + \Delta t)$. Here, $R_D$ marks 
 the constant desorption rate. 
 The probability to find the dimer attached at arbitrary time $t'$ is hence given 
 through 
 \begin{equation}
  P_\mathrm{on}(t') =e^{-t'R_D}.
\end{equation}
The signal contains weighted contributions and is obtained from 
\begin{eqnarray}\label{eq:2}
  \langle S(\tau) \rangle& = R_D \int_0^\infty P_\mathrm{on}(t') S(\tau, t')dt' \\ \nonumber
  &  = P_\mathrm{on}( \tau )S(\tau,\tau) + R_D \int_0^\tau P_\mathrm{on}(t') S(\tau, t')dt' .
\end{eqnarray}
In the second line we have used that for $t'\gtrsim \tau$ (desorption after decay
 of the probe pulse) the ion yield does not change. 
 In this case, the dimer is fully damped until 
 the probe pulse has passed.

 In a previous study, we made use of the state-independent 
 desorption scheme eq.~(\ref{eq:2}), see Ref.~\cite{Schle_245_2010}. 
 Although it is possible to find reasonable agreement with experimental findings, 
 it is an oversimplification 
 to not take into account the electronic state for desorption. 
 It is likely that only electronically excited dimers leave the droplet.  
 The electronic excitation implies a larger degree of distortion of the helium enviroment \cite{Stien_10119_2001}. 
 On the other hand, it is reasonable to assume that (slowly moving) ground state systems stay attached. 
 Therefore, we consider 
 an alternative desorption scheme 
 and only allow excited dimers to leave the droplet. 
 Desorption is again described in terms of a
 constant in time desorption rate $R_D$. The probability for the molecule 
 to desorb, however, is now proportional to the excitation probability $p_\mathrm{e}=\sum_{i \neq 0} p_i$, 
 such that eq.~(\ref{eq:6}) is replaced by 
 \begin{equation}\label{eq:11}
   p_\mathrm{off}(t', t' + \Delta t')=p_\mathrm{e}R_D\Delta t'.
 \end{equation}
 Note that in this scheme, the state after desorption does not contain any 
 electronic ground state component (see later). 
 Those dimers that do not desorb evolve according to the dissipative dynamics eq.~(\ref{eq:15}).
 In order to compensate for apparent loss of 
 ground state dimers, we need a third possible channel: 
 With probability $p_\mathrm{g} R_D \Delta t'=(1-p_\mathrm{e}) R_D \Delta t'$ the dimer 
 remains on the droplet and is projected onto its electronic ground state. 
 For the full averaged signal, we obtain
 \begin{eqnarray}\label{eq:8}
   \langle S_\mathrm{SD}(\tau) \rangle=  & P_\mathrm{on}( \tau ) S(\tau,\tau) +  p_\mathrm{e} R_D \int_0^{\tau} \!\!  P_\mathrm{on}(t') S_\mathrm{e}(\tau, t' ) dt' +\\ \nonumber
   & p_\mathrm{g} R_D \int_0^{\tau} \!\!  P_\mathrm{on}(t') S_{\mathrm{g}} (\tau, t' ) dt' . 
 \end{eqnarray}
$S_\mathrm{e}(\tau, t' )$ is the signal obtained upon 
removal of the helium influence at time $t'\lesssim \tau$ and subsequent projection on the 
excited superposition of electronic state. Likewise, $S_{\mathrm{g}}(\tau, t' )$
 is obtained upon projection on the ground state. Note that a renormalization 
 of the full wavefunction $|\Psi \rangle $ is required after projection, such that 
$\sum p_i=1$ is always valid. 

In fig.~\ref{fig:FT_833nm_SD}(b), we show the spectrogram 
for $\langle S_{SD}(\tau) \rangle$ at $\lambda=833\mathrm{\, nm}$ (scheme I). 
We use a desorption rate $R_D=0.1/\mathrm{\,ps}$, while the damping/shift parameters are not changed (as before
, $\gamma=0.15/\mathrm{ps}$ and $\Delta_{2}=-50\mathrm {cm}^{-1}$). 
 In the spectrogram, a (small) frequency shift is $\omega_A \rightarrow \omega_A'$ is observable. 
 The shift is mainly due to contributions of vibrationally damped dimers by means of the first term in eq.~(\ref{eq:8}). 
 We obtain a frequency shift $\omega_A \rightarrow \omega_A'$ in the model, but the shift 
 is more pronounced in the experiment \cite{Claas_1151_2006}. 
 The second term in eq.~(\ref{eq:2}) marks contributions from dimers which 
 are damped up to time $t'\lesssim \tau$, but are not damped afterwards, 
 i.\,e.\,upon release from the droplet. Early 
 desorbing dimers are not vibrationally relaxed and 
 the WP in the $\mathrm{\,A}\, ^1\Sigma^+_u$ state continous to reach the initial FC region. 
 From this point of view, undamped dimers may well contribute to the ion yield. 
   Indeed, contributions  to the ion yield 
   at later delay times $t \gtrsim 20\mathrm{\, ps}$ are exclusively attributed to these undamped dimers. 
   There, the oscillation frequency is near the initial gas phase value $\omega_A=\omega_{\bar{v},\bar{v}+1}$.
   
   As a final note, we find that the observed frequency shift $\omega_A \rightarrow \omega_A'$
   becomes negligible if we 
   do not include the electronic shift $\Delta_2$ in the model. 
 
  As the laser frequency is further increased,
  the vibrational energy of the WP in the $\mathrm{\,A}\, ^1\Sigma^+_u$ state increases (scheme II). Due to the larger elongation and faster dynamics, one 
  may think of larger damping of vibrational motion and/or faster desorption of dimers 
  off the droplet. In fig.~\ref{fig:PP_FT_SD_800nm}(b) we show the spectrogram at $\lambda=800\mathrm{\, nm}$, 
  as obtained for $\gamma_i=0.15/\mathrm{ps}$ and $R_D=0.5/\mathrm{ps}$, i.~e.~we assume a 
  faster desorption as before for scheme I.
  In fact, this value for the desorption rate means that dimers quickly
  leave the droplet after the pulse excitation. 
  Until desorption, electronically excited dimers are fully damped. 
  During that time, they approach an intermediate transition region with smaller 
  overlap with higher electronic states, see fig.~\ref{fig:WPs_833nm}(c). 
  As a consequence, the component $\omega_A$ fades away after several picoseconds. 
  Also, vibrational WPs of desorbed dimers do not reach 
  the FC window at the equilibrium distance, which is 
  also marked in fig.~\ref{fig:WPs_833nm}(c). 
  Therefore, an increase of the component $\omega_A$ is excluded, compare 
  fig.~\ref{fig:PP_FT_SD_800nm}(a) and (b). 
  To conclude, damping 
  in connection with fast desorption of dimers 
  explains the experimental result, see fig.~\ref{fig:FT_800nm_gphase_exp}(b), where 
  the component $\omega_A$ is only 
  visible in the beginning of the measurement and then disappears. 
  Note that dimers which remain on the droplet in the ground state, 
  are fully damped. 
  Therefore, the ground state component $\omega_X$ decreases in this model. 
\subsection{Undamped ground state WPs}\label{sec:undamp-ground-state}

Agreement with experiment can be further improved, if one 
allows for undamped motion of the vibrational WP in the ground state. 
This is of particular relevance for excitation scheme II, for which 
the result upon leaving the model parameters for damping, desorption 
and shift unchanged, but setting $\gamma_0=0$, is shown in fig.~\ref{fig:PP_FT_SD_800nm}(c). 
Frequency components at later delay times are 
exclusively attributed to the ground state motion, i.~e.~
the component $\omega_X$ is clearly visible. 
    Undamped vibrational wave packet motion 
    has been discussed 
    in terms of a critical Landau velocity $v_\mathrm{crit}$ in Ref.\cite{Schle_245_2010}. 
    The existence of $v_\mathrm{crit}$ in the 
    superfluid nanodroplet may allow for frictionless motion 
    of slowly moving ground state WPs. 
    In fig.~\ref{fig:PP_FT_SD_800nm}(c) 
    the ground state motion is assumed frictionless; 
    good agreement with the experimental result, see fig.~\ref{fig:FT_800nm_gphase_exp}(b), 
    is also obtained for nearly frictionless motion $\gamma_0 \approx 0$. 

    Note that for the observation of the 
    ground state WP, the electronic shift $\Delta_2$ is less important. 
    It only leads to a slightly different ratio between the 
     ground-and excited state frequency component.

\section{\label{sec:conclusion}Conclusions}
We consider vibrational wave packet dynamics 
of dimers attached to He nanodroplets. 
It is found that (calculated) gas phase spectra 
and spectra from dimers attached to He droplets are markedly different. 
The interaction between droplet and dimer 
influences the vibrational dimer dynamics in three ways:
Shifts, 
damping, and desorption. 
All three ingredients are taken into account in a phenomenological manner and 
each contributes to characteristic changes of the resulting spectra. 

First, we study electronic decoherence, occurring for 
instance due to slightly fluctuating shifts of electronic surfaces. 
 We find that, indeed, resulting spectra do not 
 show electronic interferences such that 
 contributions from several 
 electronic states, in particular the ground state, are 
 clearly resolved. In this way, the modeled spectrum
 is already similar to the experimental finding.

 However, pure electronic decoherence cannot explain 
 a decreasing contribution to the signal from wave packets in the first excited $\mathrm{\,A}\, ^1\Sigma^+_u$ state. 
 Consequently, vibrational damping of wave packets is taken into account, which improves 
 agreement with experiment for short delay times. 
 We find that damping is not present over the full observation timescale, 
 probably due to desorption of dimers from the droplet. 
We use a desorption scheme, which takes into account the 
occupation of electronic levels. 
 We find that the desorption rate depends on the mean vibrational 
 energy of the wave packet. At $\lambda=800\mathrm{\, nm}$, 
 desorption is fast, taking place on average several picosecond after excitation ($R_D=0.5/\mathrm{ps}$). 

 Note that in this study, all involved electronic states are spin singlet states. 
 For the WP in these singlet states, we find a relaxation rate
 which is significantly higher 
 than in previously considered spin triplet Rb$_2$ dimers attached to He droplets. 
 The smaller rate in the latter may be ascribed to the orientation 
 of the dialkali axis relative to the droplet surface. 
 Recent calculations show that the axis of spin triplet dimers 
 is oriented parallel to the droplet surface 
 while singlet states are assumed to be oriented perpendicularly \cite{Guill_6918_2011, Bovin_224903_2009}. 
 Upon the perpendicular orientation in our case, 
 the dimer might interact with the droplet more efficiently, such that relaxation is faster. 

 Damping appears to be absent for (slowly moving) wave 
 packets in the electronic ground state. 
 There, vibrational motion is found to be nearly frictionless. 
 A potential energy surface for the dimer-droplet system, which is currently underway for Rb$_2$ \cite{Guill_6918_2011}, 
 should give additional insight on dissipation rates of the wave packet. 

 The authors would like to thank Marcel Mudrich and Frank Stienkemeier for providing 
 experimental data, for fruitful discussions and 
 helpful comments. Further, we thank Alexander Eisfeld for valuable remarks.
 Support by the Deutsche
 Forschungsgemeinschaft through the research grant ``Control
 and Coherence of the Few-Particle Continuum'' is gratefully
 acknowledged. 
 Computing
 resources have been provided by the Zentrum f\"ur Informationsdienste
 und Hochleistungsrechnen (ZIH) at the TU
 Dresden. M.~S.~is a member of the IMPRS Dresden.

\bibliographystyle{iopart-num}

\providecommand{\newblock}{}

\end{document}